\documentstyle{mn}
\newread\epsffilein    
\newif\ifepsffileok    
\newif\ifepsfbbfound   
\newif\ifepsfverbose   
\newdimen\epsfxsize    
\newdimen\epsfysize    
\newdimen\epsftsize    
\newdimen\epsfrsize    
\newdimen\epsftmp      
\newdimen\pspoints     
\def\epsfbox#1{\global\def\epsfllx{72}\global\def\epsflly{72}%
   \global\def\epsfurx{540}\global\def\epsfury{720}%
   \def\lbracket{[}\def\testit{#1}\ifx\testit\lbracket
   \let\next=\epsfgetlitbb\else\let\next=\epsfnormal\fi\next{#1}}%
\def\epsfgetlitbb#1#2 #3 #4 #5]#6{\epsfgrab #2 #3 #4 #5 .\\%
   \epsfsetgraph{#6}}%
\def\epsfnormal#1{\epsfgetbb{#1}\epsfsetgraph{#1}}%
\def\epsfgetbb#1{%
\openin\epsffilein=#1
\ifeof\epsffilein\errmessage{I couldn't open #1, will ignore it}\else
   {\epsffileoktrue \chardef\other=12
    \def\do##1{\catcode`##1=\other}\dospecials \catcode`\ =10
    \loop
       \read\epsffilein to \epsffileline
       \ifeof\epsffilein\epsffileokfalse\else
          \expandafter\epsfaux\epsffileline:. \\%
       \fi
   \ifepsffileok\repeat
   \ifepsfbbfound\else
    \ifepsfverbose\message{No bounding box comment in #1; using defaults}\fi\fi
   }\closein\epsffilein\fi}%
\def\epsfsetgraph#1{%
   \epsfrsize=\epsfury\pspoints
   \advance\epsfrsize by-\epsflly\pspoints
   \epsftsize=\epsfurx\pspoints
   \advance\epsftsize by-\epsfllx\pspoints
   \epsfxsize\epsfsize\epsftsize\epsfrsize
   \ifnum\epsfxsize=0 \ifnum\epsfysize=0
      \epsfxsize=\epsftsize \epsfysize=\epsfrsize
     \else\epsftmp=\epsftsize \divide\epsftmp\epsfrsize
       \epsfxsize=\epsfysize \multiply\epsfxsize\epsftmp
       \multiply\epsftmp\epsfrsize \advance\epsftsize-\epsftmp
       \epsftmp=\epsfysize
       \loop \advance\epsftsize\epsftsize \divide\epsftmp 2
       \ifnum\epsftmp>0
          \ifnum\epsftsize<\epsfrsize\else
             \advance\epsftsize-\epsfrsize \advance\epsfxsize\epsftmp \fi
       \repeat
     \fi
   \else\epsftmp=\epsfrsize \divide\epsftmp\epsftsize
     \epsfysize=\epsfxsize \multiply\epsfysize\epsftmp   
     \multiply\epsftmp\epsftsize \advance\epsfrsize-\epsftmp
     \epsftmp=\epsfxsize
     \loop \advance\epsfrsize\epsfrsize \divide\epsftmp 2
     \ifnum\epsftmp>0
        \ifnum\epsfrsize<\epsftsize\else
           \advance\epsfrsize-\epsftsize \advance\epsfysize\epsftmp \fi
     \repeat     
   \fi
   \ifepsfverbose\message{#1: width=\the\epsfxsize, height=\the\epsfysize}\fi
   \epsftmp=10\epsfxsize \divide\epsftmp\pspoints
   \newcount\figskipcount
      \message{#1 \the\epsfysize  }
   \vbox to\epsfysize{\vfil\hbox to\epsfxsize{%
      \includegraphics{#1}%
      \hfil}}%
\epsfxsize=0pt\epsfysize=0pt}%

{\catcode`\%=12 \global\let\epsfpercent=
\long\def\epsfaux#1#2:#3\\{\ifx#1\epsfpercent
   \def\testit{#2}\ifx\testit\epsfbblit
      \epsfgrab #3 . . . \\%
      \epsffileokfalse
      \global\epsfbbfoundtrue
   \fi\else\ifx#1\par\else\epsffileokfalse\fi\fi}%
\def\epsfgrab #1 #2 #3 #4 #5\\{%
   \global\def\epsfllx{#1}\ifx\epsfllx\empty
      \epsfgrab #2 #3 #4 #5 .\\\else
   \global\def\epsflly{#2}%
   \global\def\epsfurx{#3}\global\def\epsfury{#4}\fi}%
\def\epsfsize#1#2{\epsfxsize}
\def\figinsert#1#2{\epsfbox{#1} \message{#2} }    
\def \Mpc {~h^{-1}~{\rm Mpc} }

\def \Om {\Omega_0 }

\def \bj {b_{\rm J}}

\def \kms {\rm ~km~s^{-1}}

\title[Statistical lensing of QSOs]
      {Statistical lensing of faint QSOs by galaxy clusters.}
\author[S.M. Croom \&  T. Shanks]
       {S.M.~Croom and T.~Shanks\\
        Physics Department, University of Durham, South Road, Durham, DH1 3LE,
England.}
\begin{document}

\maketitle

\begin{abstract}

We investigate  the anti-correlation  between faint high redshift QSOs and
low-redshift galaxy groups found by Boyle, Fong \& Shanks (1988), on the
assumption that it is caused by gravitational lensing of a flat
QSO number count, rather than by dust in the galaxy groups, or any
other systematic effect.  Using an 
isothermal sphere lens model, the required velocity dispersion is
$\sigma=1286^{+72}_{-91}\kms$.  With an isothermal sphere plus uniform
density plane, the velocity dispersion is $\sigma=1143^{+109}_{-153}\kms$,
while the plane density is $\Sigma_{\rm c}=0.081\pm0.032h~{\rm
g~cm}^{-2}$.  Both these values for the velocity dispersion are
considerably larger than the $\sim400-600\kms$ expected for  poor clusters
and groups and imply that the mass associated with such groups is
$\sim$4$\times$ larger than inferred from virial analyses.  If due to
lensing, this measurement clearly tends to favour high values of
$\Omega_0$.  We 
demonstrate how an estimate of $\Omega_0$ may be obtained, finding the
relation $\Omega_0=1.3(n/3\times10^{-4}h^3{\rm
Mpc}^{-3})(r/1\Mpc)(\sigma/1286\kms)^2$ where $r$ is the extent of the
anti-correlation and $n$ is the space density of groups.  In the
current data systematic errors in the determination of $n$ and $r$
may dominate this measurement, but this will be a potential
route to estimating $\Omega_0$ in improved
galaxy-QSO datasets where these systematics can be better controlled.

We have compared our result with that of Williams \& Irwin (1998) who
find a  positive correlation between bright LBQS QSOs and APM galaxies.
Because the QSO number counts are steeper at bright magnitudes, there is no
contradiction between this result and our own. Indeed, adapting the
lensing analysis of Williams and Irwin to our use of groups rather
than galaxies, we find that there is good agreement between the amplitude
of the positive cross-correlation found for the bright QSOs and the
amplitude of the negative cross-correlation found for the faint QSOs.
This analysis leads to a common estimate of $\Omega_0\sigma_8\sim3-4$.
This, however, is significantly higher than indicated from several 
other analyses. Further tests of the accuracy of the galaxy-QSO
cross-correlation results and thus their implications for $\Omega_0$ 
and $\sigma_8$ will soon  be available from the new 2dF QSO catalogue.

\end{abstract}

\section{Introduction}

Gravitational lensing by galaxies and clusters produces two different
effects in QSO surveys. At bright magnitudes, where QSO
counts are steep, a positive correlation of QSOs and foreground galaxies
or clusters can be produced, as objects intrinsically fainter than
the magnitude limit are amplified and hence artificially added to the
sample \cite{gg74}.  At fainter magnitudes, where the QSO number count
slope is much flatter, it is the reduction of observed area behind the
foreground lenses which dominates, producing a
deficit in the  background QSO number count \cite{w94}. 

Here we interpret the faint QSO-galaxy group anti-correlation result of
Boyle, Fong \& Shanks (1988)\nocite{bfs88} (hereafter BFS88) 
(see also Shanks et al. 1983 \nocite{sfgcs83}and Boyle 1986) in
terms of gravitational lensing. 
This result was first interpreted in terms of dust in foreground
galaxies and clusters obscuring background QSOs. However, observations
of galaxy groups and clusters do not show significant
amounts of dust \cite{ferguson93} and the limits are at a level which make
the dust hypothesis uncomfortable. Previously Rodrigues-Williams and
Hogan (1994) have suggested that the anti-correlation result may be due
to lensing and this is the avenue we shall pursue here. The results in
this paper are based in part on those of Croom (1997). 

If correct, the lensing hypothesis would allow important constraints to
be placed on cosmology and large-scale structure.  The deficit of QSOs
near a group or cluster can be used to weigh that structure. This
method has the 
advantage over other mass estimates in that it allows a  measurement of
the absolute mass of the cluster, while other estimators such as the
measurement of velocity dispersions and the observation of shear due to
strong lensing are effectively measuring the gradient of the cluster
potential \cite{btp95}. Other authors (e.g. Taylor et
al. 1998)\nocite{tdbbv98} have looked 
for a deficit of galaxies behind foreground clusters to measure the
lensing magnification. The advantage of using QSOs over galaxies is that
they are easier to distinguish as background objects and their redshift
distribution is well known.  Of course, there is the
disadvantage that QSOs are rare objects and so cannot be used
to examine the mass distribution of individual clusters, however they
can be used to investigate the properties of a distribution of clusters.

Previous searches for QSO-galaxy correlations at brighter magnitudes have
produced varying results with most showing the observational
evidence for QSO-galaxy associations, (see Table 1 of Wu
1994)\nocite{w94} but the statistical basis for most of the results was
limited. Recently, Williams \& Irwin (1998) have found a strong
positive  correlation between $\sim60$ $B<18$ LBQS \cite{lbqs95} $z>1$
QSOs and APM galaxies.  Below
we shall compare their results with ours to check whether these two
observations provide a consistent picture for the mass distribution in
the Universe. 

Section 2 reviews the lensing model we use in this paper.  In
Section 3 we compare these models to the BFS88 data. In Section 4 we
compare the results of Williams \& Irwin with those of BFS88. We present
our conclusions in Section 5.

\section{Statistical Gravitational Lensing}\label{lens_theory}

We use two analytic mass profiles to fit the observed
anti-correlation; the first, and simplest, of
these being the single isothermal sphere (SIS), which 
gives a gravitational lensing amplification of 
\begin{equation}
A=\frac{\theta}{\theta-\theta_{\rm E}},
\,\,\,\,\,\,\,\theta>\theta_{\rm E},
\label{sis}
\end{equation}
(e.g. Wu 1994) where $\theta_{\rm E}$ is the Einstein radius, the
radius within which multiple images can occur.  For the SIS case this is
\begin{equation}
\theta_{\rm E}=4\pi\frac{D_{\rm ls}}{D_{\rm
s}}\left(\frac{\sigma}{c}\right)^{2},
\label{thetaeeq}
\end{equation}
where $D_{\rm s}$, $D_{\rm l}$ and $D_{\rm ls}$ are the angular
diameter distances from the observer to the source, the observer to
the lens and the lens to the source, respectively.
In our second mass profile we add a uniform density plane to
the isothermal profile (SIS+plane).  This could be a good approximation to
the effects of 
clustering and large scale structure (as pointed out by Wu et al.,
1996)\nocite{wfzq96}, because a distribution of isothermal spheres
with an
auto-correlation function of the form $\xi(r)\sim r^{-2}$ produces
a uniform mass surface density.  The globally measured auto-correlation
function slope is $\sim-1.8$ \cite{dmsdy88}, which produces a sheet of matter which
is uniform to better than $10\%$ on the scales of interest.  The
amplification then becomes
\begin{equation}
A=\frac{\theta}{\theta-\theta_{\rm E}/(1-\Sigma_{\rm c}/\Sigma_{\rm
crit})}\frac{1}{(1-\Sigma_{\rm c}/\Sigma_{\rm crit})^2},
\label{sisplane}
\end{equation}
(e.g. Wu et al. 1996) where $\Sigma_{\rm c}$ is the mass surface density in the plane and
the critical surface density, $\Sigma_{\rm crit}$, is
\begin{equation}
\Sigma_{\rm crit}=\frac{D_{\rm s} c^2}{D_{\rm ls}D_{\rm l} 4\pi G}.
\end{equation}

Gravitational lensing can cause an over- or under-density of source objects
near to the lens.  The ratio of observed surface density to the true
surface density (unlensed) is the enhancement factor, $q$, given by
\begin{equation}
q=\frac{N(<m+2.5\log(A))}{N(<m)}\frac{1}{A}
\end{equation}
\cite{n89}, where $A$ is the amplification factor.  $N(<m)$ is the
integrated number count
of source objects brighter than magnitude $m$.  We note that $q$
depends on the source counts fainter than the limit of the survey.
With a number count of the form $N(<m)\propto10^{\alpha m}$, we then
find an angular cross-correlation function $\omega_{\rm CQ}(\theta)$
that is described by
\begin{equation}
\omega_{\rm CQ}(\theta)=q-1=A^{2.5\alpha-1}-1.
\label{omega_a}
\end{equation}

\section{The Correlation of Durham/AAT QSOs and Galaxy Groups}

We look at the result from the Durham/AAT UVX Survey \cite{boyle86,bfsp90}
which shows an anti-correlation between UVX QSO candidates and galaxy
groups (BFS88).  This cross-correlation was carried out within 7 UKST
fields, using COSMOS scans of photographic plates.  Spectroscopy of
the UVX catalogue \cite{bfsp87} suggested that with a
colour limit of $u-b<-0.4$ there was $\sim55$ per cent contamination by
Galactic stars.  In the BFS88 analysis the UVX criterion
was tightened to $u-b<-0.5$, reducing contamination to 25 per cent
while keeping 85 per cent of the QSOs.  The
UVX catalogue was then split into two magnitude limited samples,
$17.9<b<19.9$ and $17.9<b<20.65$.  The galaxy catalogue consists of all
galaxies to a limit of $\bj=20.0$ and the cluster sample was created
using a `friends-of-friends' algorithm \cite{gt77,sfs88}.  Groups of seven
or more galaxies with density greater than 8 times the average for the
field were classed as clusters, which amounted to 10 per cent of the
total number of galaxies.  BFS88 performed a cross-correlation analysis
between the entire galaxy catalogue and the UVX sample but no
significant correlation
was found on any scale.  Cross-correlation of cluster galaxies with the
UVX catalogue resulted in negative correlations on scales $<10'$ for
both  samples, the brighter sample showing a marginally more negative
clustering signal.  This can be interpreted as a decrease in
contamination from smaller photometric errors in the brighter
sample, but a second effect is that the QSO $N(m)$ slope will be
steeper at this brighter limit, thus a smaller anti-correlation might
be expected.  Given that our results are sensitive to the exact
shape and position of the break, we restrict our analysis to the fainter
sample, with the proviso that in new larger samples the
anti-correlation as a function of magnitude will provide an important
test of the gravitational lensing hypothesis.

\subsection{A comparison of lensing models and the data}

\begin{figure}
\centering
\centerline{\epsfxsize=8.5truecm \figinsert{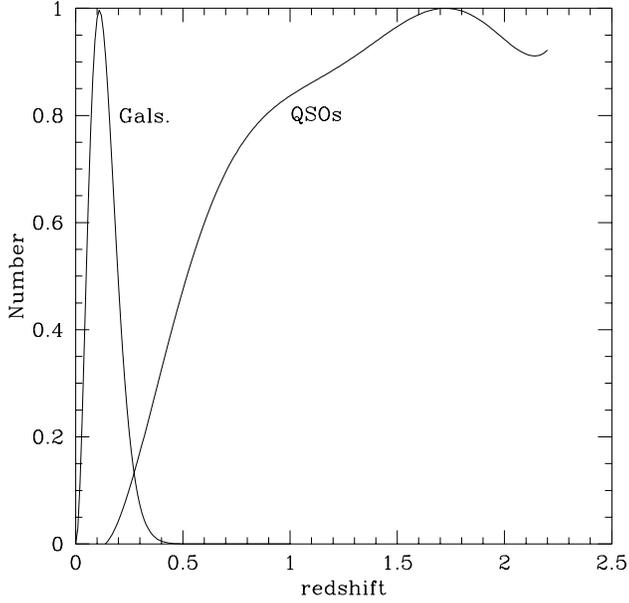}{0.0pt}}
\caption{Redshift distributions assumed for the galaxy and QSO
populations in the lensing models described in the text.  The
normalizations of the two curves are arbitrary.}\label{lensnz}
\end{figure}

The effect of gravitational lensing is strongly dependent on the slope
of the QSO number-magnitude relation at faint magnitudes.  We use the
number counts from Boyle, Shanks \& Peterson (1988)\nocite{bsp88}
which give an asymptotic faint end slope of $\sim0.28$.  We have
confirmed that this is a reasonable representation of the integral QSO number
count at $\sim1$ mag fainter than our magnitude limit, the region from
where we expect amplified QSOs to come, by using the deeper data of
Boyle, Jones \& Shanks (1991).  A flatter
slope would clearly reduce the lensing mass required.  The separation of
observer, lens and source also affects the lensing amplification.  To
take this into account in our model, we integrate the known QSO
redshift distribution over the effective range of the Durham/AAT survey
($0.3<z<2.2$).  This gives us an effective lensing amplification for a
particular lens mass.  For the galaxies we assume the analytic form of 
$N(z)$ given by Baugh \& Efstathiou 
(1993)\nocite{be93}:
\begin{equation}
\frac{{\rm d}N}{{\rm d}z}\propto z^2{\rm
exp}\left[-\left(\frac{z}{z_{\rm c}(m)}\right)^{3/2}\right],
\label{nzeq}
\end{equation}
where $z_{\rm c}=(0.016(b_{\rm J}-17.0)^{1.5}+0.046)/1.412$.  This is
shown integrated to $b_{\rm J}=20$ in Fig. \ref{lensnz} along with a
polynomial fit to the QSO $N(z)$ \cite{sb94}.  The
two populations occupy almost completely independent volumes, less
than $1\%$ of the QSOs are at $z<0.3$ while less than $0.5\%$ of the
galaxies are at $z>0.3$.  We assume an $\Om=1$ cosmology throughout
this analysis, but it should be noted that when the lensing mass is at
low redshift ($z\sim0.1$) cosmology has a relatively small effect as
$D_{\rm s}\sim D_{\rm ls}$.

\begin{figure}
\centering
\centerline{\epsfxsize=9.0truecm \figinsert{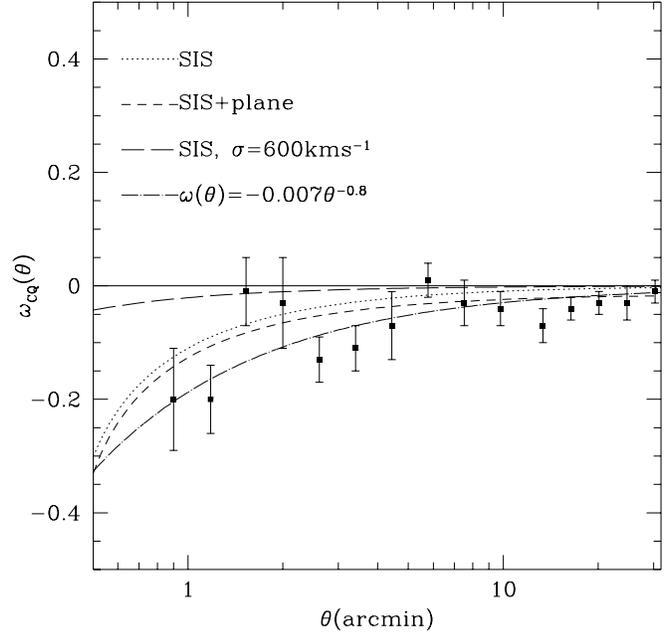}{0.0pt}}
\caption{UVX-cluster cross-correlation for the faint UVX sample
of BFS88 showing the best fit models; the dotted line shows the best
fit SIS  model, while the
dashed line shows the SIS+plane model.  The
long-dashed line shows a lensing SIS model with $\sigma=600\kms$.
Also shown as the dot-dash line is the best fit $\theta^{-0.8}$ power law.}
\label{datamod}
\end{figure}

\begin{figure}
\centering
\centerline{\epsfxsize=9.0truecm \figinsert{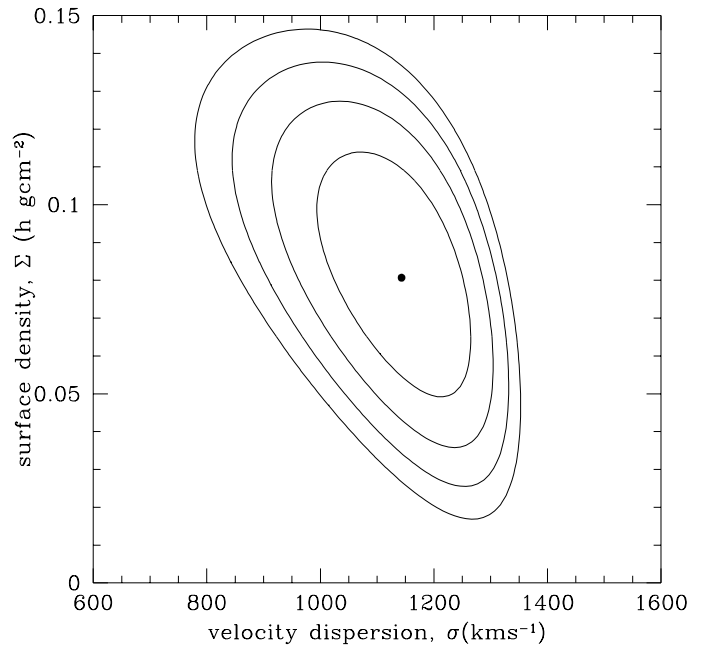}{0.0pt}}
\caption{Confidence contours for the fit of the two component
SIS+plane model to the cross-correlation result for the faint UVX
sample, corrected for $25\%$ stellar
contamination in the QSO survey.  Contours are at
$\Delta\chi^2=1,2,3,4$ ($\Delta\chi^2=1$ and $4$ correspond to $1$ and
$2$ $\sigma$ errors respectively).}\label{conf} 
\end{figure}

We compare the SIS lens model (from Eqs. \ref{sis} and \ref{omega_a})
to the cross-correlation result $\omega_{CQ}(\theta)$ of the faint
sample shown in Fig. \ref{datamod}.  We  have allowed
for 25\% contamination of the QSOs by randomly distributed stars.
Using a minimum $\chi^2$ fit the velocity dispersion is
$\sigma=1286^{+72}_{-91}\kms$ (reduced $\chi^2=1.44$).  The dotted
line in Fig. 
\ref{datamod} shows this model.  The SIS+plane model (Eqs.
\ref{sisplane} and \ref{omega_a}) is shown as the dashed line in Fig.
\ref{datamod}, the best fit in this case has velocity dispersion
$\sigma=1143^{+109}_{-153}\kms$ and the surface density in the plane of
$\Sigma_{\rm c}=0.081\pm0.032h~{\rm 
g~cm}^{-2}$.  The reduced $\chi^2$ for this fit is $1.07$.  The confidence
levels for this fit are shown in Fig. \ref{conf}. 

The values found are significantly larger than directly measured
velocity dispersions of poor clusters and groups.  However, because
BFS88 correlate cluster members with UVX objects, they effectively
weight each cluster by the number of member galaxies.  Stevenson et
al., (1988)\nocite{sfs88} find the fraction of clusters as a function
of the number of members
follows an approximate power-law with a slope of $-2.2$.  From this
we can calculate the mean cluster membership, $\bar{n}$.  Integrating
the relationship between $n=7$ and $n=50$ gives $\bar{n}=15$.
However, a member weighted mean gives $\bar{n}=20$.  Thus, the BFS88
result is probing
clusters which typically have $\sim20$ members.  The density on the
sky of these clusters is $\sim0.8$ deg$^{-2}$, which can be compared
to the density of clusters in the APM Cluster Catalogue \cite{dalton97}
of $\sim0.2$ deg$^{-2}$ and the density of richness class 0 or
greater Abell clusters which is $\sim0.1$ deg$^{-2}$ \cite{aco89}.  Thus
an `average' cluster used by BFS88 is significantly poorer than Abell
richness clusters. The velocity dispersion that might be expected for
clusters of this richness is $\sigma\sim400-600\kms$ (e.g. Ratcliffe
et al. 1998).  For
comparison, a lensing model corresponding to a velocity dispersion of
$ 600\kms$ is shown in Fig. \ref{datamod}; the model is  formally
rejected at $>5\sigma$. It therefore appears that the masses implied for the
galaxy groups from lensing are  $\sim4$ times bigger than expected from
virial analyses. 

Although  the addition of the uniform plane to mimic the effects of
clustering of clusters helps to improve the fit to $\omega_{\rm QC}$, Fig.
\ref{conf} shows  that this only reduces the velocity dispersion of the clusters
for  high values of $\Sigma_{\rm c}$. Wu et al., (1996)\nocite{wfzq96}
find that the maximum mass likely to be associated with lenses from
large-scale structure is $\sim0.01-0.02h~{\rm g~cm}^{-2}$, which assumes
that the matter density in clusters is near the critical density (i.e.
$\Omega_{\rm clus}\sim1$).  Values of  $\Sigma_{\rm c}$ in this range
are only compatible with group velocity dispersions greater than
1000$\kms$ (see Fig. \ref{conf})

We now demonstrate how the lensing estimates of average group mass, via the
velocity dispersion $\sigma$, could be used to obtain a new estimate
of $\Omega_0$. Using
the 0.8deg$^{-2}$ sky density of groups from above, we infer an
approximate space density of galaxy groups in the range $\rm n = 
2-4\times10^{-4} h^3 Mpc^{-3}$. The lower value comes from integrating the
proper volume to z=0.1 and assuming all groups are detected to this
redshift, and on the basis of  Fig. \ref{lensnz} that this contains  half
the group sky density. The higher value comes from using the galaxy n(z)
in Fig. \ref{lensnz} to derive the galaxy selection function.  This is
multiplied by the proper volume and the product is integrated to
z=0.7, to give the effective volume from which the group space density
is then derived.  Multiplying by the estimated mass per group obtained by
integrating the isothermal sphere profile out to a radius $r$ leads to
the estimate of $\Omega_0$.  We
therefore find  \begin{equation}
\Omega_0=1.3 \left(\frac{n}{0.0003h^3 {\rm Mpc}^{-3}}\right)
\left(\frac{r}{1h^{-1}{\rm Mpc}}\right)
\left(\frac{\sigma}{1286\kms}\right)^2,
\end{equation}
where $r$ is now the extent of the anti-correlation, or the effective
extent of the isothermal sphere.  We note that the
dependence on $h$ cancels out and there is no dependence on any biasing
parameter. The above scaling values represent our best estimate for $n$, $r$
and $\sigma$. The value for r is obtained from consideration of  Fig.
\ref{datamod} where $\theta \sim 10'$ corresponds to 1$\Mpc$.
The errors on $n$ and $r$ are unfortunately likely to be dominated by
systematic components, with each potentially varying by a factor of
$2$.  This is due to the approximate methods used to determine both the
surface density of clusters, the space density of clusters, and the
limiting scale of the anti-correlation.  We also note that the
measurement of $\sigma$ is dependent on $\Omega_0$ through the angular
diameter distance terms in eq. \ref{thetaeeq}, although at the
redshifts considered this effect is small (a $\sim25\%$ difference in
mass between $\Omega_0=1$ and $\Omega_0=0$), and may currently be
dominated by more serious systematic effects.  More meaningful error
estimates must await larger galaxy-QSO datasets, possibly with full redshift
information for the galaxies as well as the QSOs, where the extent of the
anti-correlation and the group density can be better defined.

A further final problem is that this analysis for $\Omega_0$ also assumes that
the groups are physically real and not significantly contaminated by accidental
line-of-sight over-densities on the sky, which Stevenson et al (1988) suggested
was a possibility. Of course, even accidental over-densities will act as lenses
and it is not yet clear how sensitive this estimate of $\Omega_0$ is to
this type of contamination. Again, in a survey which also has full galaxy
redshift information, such as the forthcoming 2dF QSO/galaxy redshift survey,
the effects of spurious groups could be more easily
checked.
  
\section{Comparison with the LBQS-Galaxy Cross-Correlation Result }

We now compare our conclusions with those of Williams \& Irwin (1998)
(henceforth WI98)
\nocite{wi98} who have found a strong positive correlation between APM
galaxies and LBQS QSOs which is significant out to scales of $\sim60'$.
These authors find that their positive correlation is an order
of magnitude larger than that expected from a model with $\Omega_0=0.3$
and galaxy bias of $b\sim1$,  based on a comparison of the
galaxy-galaxy angular correlation function and the QSO-galaxy
cross-correlation function.  WI98 derive the relation:

\begin{equation}
\omega_{\rm QG}(\theta)\simeq(2\tau/b)(2.5\alpha-1)\omega_{\rm GG}(\theta).
\end{equation}

Here $\alpha$ is the slope of the QSO number counts, $b$ is the galaxy
bias, assumed to be constant as a function of scale and $\tau$ is the
optical depth of the lenses:

\begin{equation}
\tau=\rho_{\rm crit}\Omega_0\int^{z_{\rm max}}_{0}\frac{(c{\rm
d}t/{\rm d}z)(1+z)^3}{\Sigma_{\rm crit}(z,z_{\rm s})}{\rm d}z.
\label{tau}
\end{equation}

For a given value of $\Omega_0$, a  bias value can therefore be
found. This analysis can easily  be applied to the faint QSO
anti-correlation result of  BFS88.  We fit a power law with a slope of
--0.8 to the auto-correlation of clusters measured by Stevenson et
al. (1988), finding $\omega_{\rm
CC}(\theta)=(0.140\pm0.053)\theta^{-0.8}$.  We then also fit a --0.8
power law to the anti-correlation between QSOs and clusters, which we
find to be $\omega_{\rm CQ}(\theta)=(-0.0071\pm0.0059)\theta^{-0.8}$.
With an assumed number count slope of $0.28\pm0.02$ these values then
imply a value of $\tau/b=0.085\pm0.077$.  We assume $\Omega_0=1$ and
integrate Eq. \ref{tau} to $z_{\rm max}=0.2$, the redshift at which
the $N(z)$ relation described by Eq. \ref{nzeq} falls to half its peak
value, this gives $\tau=0.021$.  We therefore find a bias value of the
clusters used in this analysis of $b_{\rm C}=0.25\pm0.23$.  If we use
an $\Omega_0=0.3$ model, as used by WI98, then $\tau$ and
therefore $b_{\rm C}$ will fall by a factor of $\sim3$.  We should
note here that the errors are large, and there is some uncertainty in
this procedure; if we integrate $\tau$ to where the $N(z)$ drops to $3/4$
($z=0.16$) or $1/4$ ($z=0.24$) of its peak value we find $\tau=0.015$
and $0.029$ respectively.  However even if we take the largest
reasonable value of $\tau$, then $b_{\rm C}=0.34\pm0.031$, which is
still an order of magnitude lower than expected for clusters.
Thus, in 
rough agreement with $b_{\rm G}\sim0.07$ from WI98, we
find a bias value estimated 
from statistical lensing which is an order of magnitude less than
predicted by other methods.  We also note that the WI98
result is consistent with the QSO-galaxy cross-correlation
measured by BFS88, although BFS88 do not find a significant
anti-correlation.

Although our result appears to be consistent with WI98,
they are both clearly significantly out of line with other current
estimates of the combination of $\Omega_0$ and bias.  The
space density of galaxy clusters gives $\Omega_0^{0.5}\sigma_8\simeq0.5$
(Eke et al. 1998) and dynamical estimates such as the measurement of
redshift space distortions give similar values (e.g.  Ratcliffe et
al. 1998).  We could possibly appeal
to the scale dependence of bias to bring these different results
into agreement, however, this would require an order of magnitude
change in bias over a scale of $\sim10\Mpc$.  Taken at face value the
above lensing result appears to suggest much more mass is present in
the Universe than is detected from the distribution and motion of
galaxies.  

\section{Discussion and Conclusions}

BFS88 originally interpreted the UVX QSO-cluster anti-correlation as
being due to absorption by dust present in clusters, the required
amount of absorption being $A_{\rm B}\simeq0.2$ mag.  Ferguson
(1993)\nocite{ferguson93} finds no evidence for any reddening due to
dust in clusters, and the 90\% upper limit on
the reddening is $E(B-V)=0.06$. This  upper limit  is just
consistent with the required absorption assuming $A_{\rm
B}=4.10E(B-V)$, and  it is therefore still possible that  lensing {\it
and} absorption could both play a part in producing the anti-correlation
result.  However, it is impossible for dust in groups to also provide an
explanation for the strong positive QSO-galaxy correlation found by
Williams \& Irwin and if their result is due to lensing then  an
anti-correlation is expected at faint QSO magnitudes which is comparable
to that discussed here.  If both results prove to be real, the
simplest  interpretation is that both are due to gravitational lensing.

Assuming that the measured anti-correlation is due to gravitational
lensing, we find that fitting an isothermal sphere model for the cluster
potentials gives a larger than expected velocity dispersion.  Adding a
uniform density plane to the mass profile does not significantly affect
this conclusion.  These lensing mass estimates suggest cluster/group
masses which are $\sim4$ times larger than expected from virial
theorem analyses.  We discuss a potential method to determine $\Omega_0$
from this type of mass estimate combined with a cluster/group space
density measurement.  We demonstrate this method with the current
data, although an accurate measure of $\Omega_0$ will have to wait for
larger and better controlled galaxy-QSO dataset.

We find consistency between the high $\Omega_0/b\sim3-4$ value
implied by the strong positive QSO-galaxy cross-correlation seen at
bright QSO magnitudes (Williams \& Irwin 1998) and the negative
QSO-galaxy cross-correlation seen at faint QSO magnitudes (BFS88),  if
lensing is assumed to cause  both effects. Applying the method of Williams
\& Irwin to both these cross-correlation results gives 
$\Omega_0\sigma_8\sim3-4$ (where $\sigma_8\sim1/b$) and the inferred values of
$\Omega_0^{0.5}\sigma_8$ are therefore 6--8 times higher than those
inferred from arguments based on the space-density of rich clusters.

Of course, it is still possible that some  combination of systematic and
random errors have contrived to produce the positive QSO-galaxy
correlation seen in LBQS and the anti-correlation detected by BFS88. The
importance of the above results suggests that it is vital to make further
observational checks as to the reality of the QSO-galaxy cross-correlation
signal.  Fortunately, extended analyses of the above type will soon be
possible  with the completion of new large redshift surveys such as the
2dF QSO Redshift Survey and the 2dF Galaxy Redshift Survey.  These two
samples with 25000 QSOs and 250000 galaxies covering the same areas of sky
should allow a definitive measurement of the cross-correlation function
between background QSOs and galaxies at both bright and faint QSO
magnitudes.

\section*{acknowledgements}

SMC acknowledges the support of a Durham University Research
Studentship.  This paper was prepared using the facilities of the
Durham {\small STARLINK} node.

{}

\end{document}